\documentclass{ws-procs9x6}

\begin{document}

\title{The Nuclear Equation of State at high densities}

\author{Christian Fuchs}

\address{Institut f\"ur Theoretische Physik, \\Universit\"at 
T\"ubingen, \\D-72076 T\"ubingen, Germany\\
E-mail: christian.fuchs@uni-tuebingen.de}

\begin{abstract}
Ab inito calculations for the nuclear many-body problem make 
predictions for the density and isospin dependence of the 
nuclear equation-of-state (EOS) far away from the saturation point 
of nuclear matter. I compare predictions 
from microscopic and phenomenological approaches. 
Constraints on the EOS derived 
from heavy ion reactions, in particular from subthreshold kaon 
production, as well as constraints from neutron stars are discussed.  
\end{abstract}

\bodymatter
\section{Introduction} 
Heavy ion reactions provide the only possibility to reach nuclear 
 matter densities beyond saturation density $\rho_0 \simeq 0.16~{\rm 
 fm}^{-3}$. Transport calculations indicate that in  
the low and intermediate energy range  
$E_{\rm lab}\sim 0.1\div 1$ AGeV nuclear densities between $2\div 3 \rho_0$  
are accessible while the highest baryon densities ($\sim 8 \rho_0$)   
will probably be reached in the energy range of  
the future GSI facility FAIR between $20\div 30$  
AGeV. At even higher  
incident energies transparency sets in and the matter becomes less baryon  
rich due to the dominance of meson production. The isospin dependence of 
the nuclear forces which is at present only little 
constrained by data will be explored by the forthcoming radioactive beam 
facilities at FAIR/GSI \cite{sis200}, SPIRAL2/GANIL and RIA \cite{ria}. 
Since the knowledge of the  
nuclear equation-of-state (EOS) at supra-normal densities and extreme 
isospin is essential for  
our understanding of the nuclear forces as well as for astrophysical  
purposes, the determination of the EOS was already one of the primary  
goals when first relativistic heavy ion beams started to operate in  
the beginning of the 80ties. In the following I 
briefly discuss the knowledge about the nuclear EOS at {\it moderate 
densities and temperatures}. For more details see e.g. Ref. \cite{WCI}.

Models which make predictions on the nuclear EOS can roughly be divided  
into three classes: {\it phenomenological density functionals} such 
as  Gogny or Skyrme forces \cite{gogny02,skyrme04,reinhard04} and  
relativistic mean field (RMF) models \cite{rmf}, 
{\it effective field theory} (EFT) and {\it ab initio approaches}. 
In EFT a systematic expansion of the  
EOS in powers of density, respectively the Fermi momentum $k_F$ is
performed. EFT
can be based on density functional theory \cite{eft1,eft2} or e.g. on  
chiral perturbation theory \cite{lutz00,finelli,weise04}.
Ab initio approaches are ased on high precision  
free space nucleon-nucleon interactions and the nuclear many-body  
problem is treated microscopically. Predictions for the nuclear EOS  
are essentially parameter free. Examples are variational  
calculations \cite{akmal98}, Brueckner-Hartree-Fock (BHF)  
\cite{zuo02,zuo04} or relativistic Dirac-Brueckner-Hartree-Fock (DBHF)  
\cite{terhaar87,boelting99,dalen04} and Greens functions  
Monte-Carlo approaches \cite{dickhoff04,carlson03}.  In the follwoing 
I will mainly concentrate on the DBHF approach.

\section{The EOS from ab inito calculations} 
\begin{figure}[h] 
\centerline{
\includegraphics[width=0.6\textwidth]{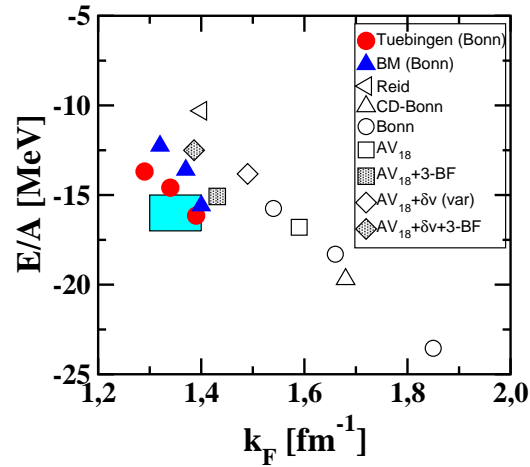}}
\caption{Nuclear matter saturation points from relativistic (full symbols)  
and non-relativistic (open symbols) Brueckner-Hartree-Fock calculations  
based on different nucleon-nucleon forces. The diamonds show results from  
variational calculations. Shaded symbols denote calculations which  
include 3-body forces. The shaded area is the empirical region of  
saturation. Figure is taken from Ref. \protect\cite{fuchs05}.
} 
\label{coester_fig} 
\end{figure} 
In  {\it ab initio} calculations based on many-body techniques one  
derives the energy functional from first principles, i.e. treating  
short-range and  many-body correlations explicitely. A typical  
example for a successful  many-body approach is Brueckner theory \cite{gammel}.  
In the relativistic Brueckner approach the nucleon  
inside the medium is dressed by the self-energy $\Sigma$.  
The in-medium T-matrix which is obtained from  
the relativistic Bethe-Salpeter (BS) equation plays the role  
of an effective two-body interaction which contains all short-range  
and many-body correlations of the ladder approximation.  
Solving the BS-equation the Pauli principle is respected  
and intermediate scattering states are projected  
out of the Fermi sea. 
The summation of the  T-matrix over the occupied states inside the Fermi sea  
yields finally the self-energy in Hartree-Fock approximation. This coupled set of  
equations states a self-consistency problem which has to be solved  
by iteration.  
 
In contrast to relativistic DBHF calculations which came up in the late 
80ties non-relativistic BHF theory has already almost half a century's 
history. The first numerical calculations for nuclear matter were carried 
out by Brueckner and Gammel in 1958 \cite{gammel}. Despite strong  
efforts invested in the development of improved solution techniques for  
the Bethe-Goldstone (BG) equation, the non-relativistic counterpart of the  
BS equation, it turned out that, although such calculations were able to 
describe the nuclear saturation mechanism qualitatively, they failed  
quantitatively. Systematic studies for a large number of NN 
interactions were always allocated on a 
so-called {\it Coester-line} in the $E/A-\rho$ plane which does not 
meet the empirical region of saturation. In particular modern  
one-boson-exchange (OBE) potentials  
lead to strong over-binding and too large saturation densities where  
relativistic calculations do a much better job.  
 
Fig. \ref{coester_fig} compares the saturation points of nuclear matter  
obtained by relativistic Dirac-Brueckner-Hartree-Fock (DBHF) calculations  
using the  Bonn potentials \cite{bonn} as bare $NN$ interactions   
to non-relativistic Brueckner-Hartree-Fock calculations for various  
 $NN$ interactions. The DBHF results are taken from Ref. \cite{bm90} (BM)  
and more recent calculations based on improved techniques are  
from Ref. \cite{boelting99} (T\"ubingen). 
\begin{figure}[h] 
\centerline{
\includegraphics[width=0.8\textwidth]{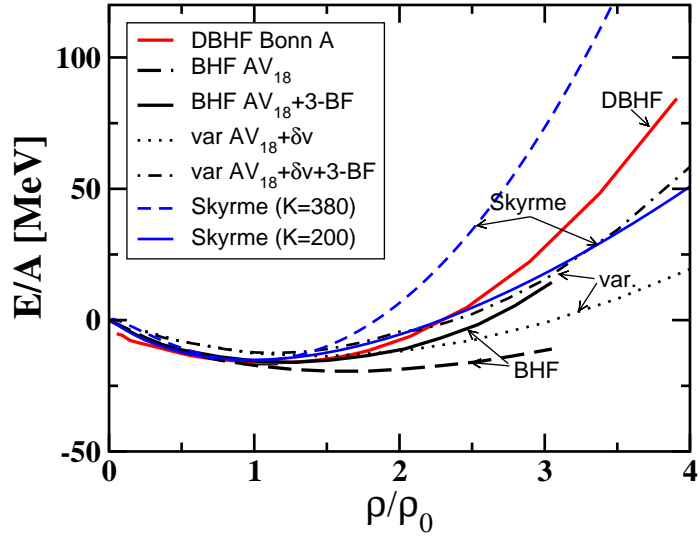}}
\caption{Predictions for the EOS of symmetric  
nuclear matter from microscopic ab initio  
calculations, i.e. relativistic DBHF \protect\cite{boelting99},  
non-relativistic BHF \protect\cite{zuo02} and variational  
\protect\cite{akmal98} calculations. For comparison also  
 soft and hard Skyrme forces are shown. 
Figure is taken from Ref. \protect\cite{fuchs05}.} 
\label{dbeos_fig} 
\end{figure} 
Several reasons have been discussed in the literature in order to 
explain the success of the relativistic treatment (see e.g. discussion 
in Ref. \cite{honnef}). 
Three-body forces (3-BFs) have extensively been studied within 
non-relativistic BHF \cite{zuo02} and variational calculations  
\cite{akmal98}. Both approaches shwon in Fig. \ref{dbeos_fig} 
are based on the latest ${\rm AV}_{18}$ version of the Argonne  
potential. The variational results shown  
contain boost corrections ($\delta v$) which account for relativistic kinematics and 
lead to additional repulsion \cite{akmal98}. 

The contributions from 3-BFs are in total   
repulsive which makes the EOS harder and non-relativistic  
calculations come close to their relativistic counterparts.  
The same effect is observed in variational calculations \cite{akmal98}  
shown in Fig. \ref{dbeos_fig}. It is often argued that in non-relativistic 
treatments 3-BFs play in some sense an equivalent role as the 
dressing of the two-body interaction by in-medium spinors in  
Dirac phenomenology. Both mechanisms lead indeed to an effective density 
dependent two-body interaction $V$ which is, however, of different 
origin. One class of 3-BFs involves virtual excitations of 
nucleon-antinucleon pairs. Such Z-graphs  are 
in net repulsive and can be considered as a renormalization of  
the meson vertices and propagators. A second class of 3-BFs is  
related to the inclusion of explicit resonance degrees of freedom.  
The most important resonance is the $\Delta$(1232)  
isobar which provides at low and intermediate  
energies large part of the intermediate range attraction.

Fig. \ref{dbeos_fig} compares the equations of state from the  
different approaches: DBHF from Ref. \cite{boelting99}   
based the Bonn A interaction\footnote{The high density behavior of the  
EOS obtained with different interaction, e.g. Bonn B or C is  
very similar. \cite{boelting99}}  \cite{bonn},  
BHF  \cite{zuo02}  
and variational calculations \cite{akmal98}. The latter ones are  
 based on the Argonne ${\rm AV}_{18}$ potential and include  
3-body forces. All the approaches use modern high precision $NN$  
interactions and represent state of the art calculations. Two   
phenomenological Skyrme functionals which correspond to the  
limiting cases of a soft (K=200 MeV) and a hard (K=380 MeV) EOS are  
shown as well. In contrast to the Skyrme  
interaction where the high density behavior is fixed by the 
parameteres which determine the compression modulus,  
in microscopic approaches the compression modulus is  
only loosely connected to the curvature at saturation density.   
DBHF Bonn A has e.g. a compressibility of K=230 MeV.  
Below $3\rho_0$ both are not too far  
from the soft Skyrme EOS. The same is true for BHF including 3-body  
forces.

When many-body calculations are performed, one has to keep in mind that 
elastic $NN$ scattering data constrain the interaction only 
up to about 400 MeV, which corresponds to the pion 
threshold. $NN$ potentials differ essentially 
in the treatment of  the short-range part. A model independent 
representation of the $NN$ interaction can be obtained in EFT approaches where 
the unresolved short distance physics is replaced  by simple contact 
terms. In the framework of chiral EFT the $NN$ interaction has been 
computed up to N$^3$LO \cite{entem03,epelbaum05}. An alternative 
approach which leads to similar results is based on 
renormalization group (RG) methods \cite{lowk03}. In the $V_{\rm low~k}$ 
approach a low-momentum potential is derived from a given 
realistic $NN$ potential by integrating out 
the high-momentum modes using RG methods. 
When applied to the nuclear many-body 
problem low momentum interactions do not require a full resummation 
of the Brueckner ladder diagrams but can already 
be treated within second-order perturbation theory \cite{bogner05}. 
However, without  repulsive three-body-forces 
isospin saturated nuclear matter was found to collapse. Including 3-BFs 
first promising results have been obtained with $V_{\rm low~ k}$ 
\cite{bogner05}, however, nuclear saturation is not yet described 
quantitativley. 
\subsection{EOS in symmetric and asymmetric nuclear matter} 
\begin{figure}[h] 
\centerline{
\includegraphics[width=0.8\textwidth]{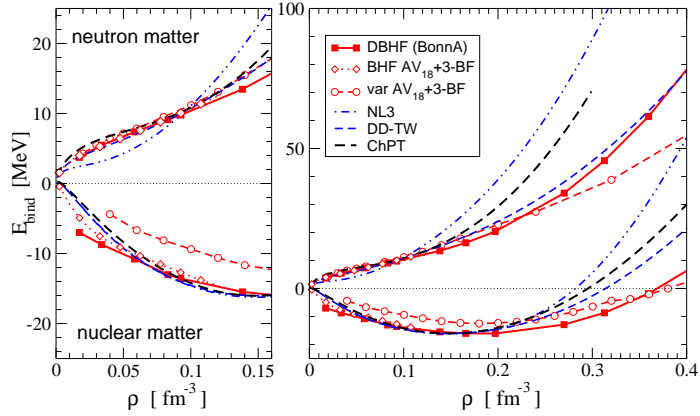}}
\caption{EOS in nuclear matter and neutron matter.  
BHF/DBHF and variational calculations are compared to  
phenomenological density functionals NL3 and DD-TW and  
ChPT+corr.. The left panel zooms the low density range. 
The Figure is taken from Ref. \protect\cite{WCI}.  
} 
\label{nmeos_fig} 
\end{figure} 
Fig. \ref{nmeos_fig} compares now the predictions for nuclear and neutron  
matter from microscopic  
many-body calculations -- DBHF \cite{dalen04}  and the 'best' 
variational calculation with 3-BFs and boost corrections  
\cite{akmal98} -- to phenomenological approaches and to EFT. As typical  
examples for relativistic functionals we take NL3 \cite{nl3} as one  
of the best RMF fits to the nuclear chart and a  
phenomenological density dependent  
RMF functional DD-TW from Ref. \cite{typel99}.  
ChPT+corr. is based on chiral pion-nucleon dynamics \cite{finelli}  
including condensate fields and fine tuning to finite nuclei. 
As expected the phenomenological functionals  
agree well at and below saturation density where they are constrained  
by finite nuclei, but start to deviate substantially at supra-normal  
densities. In neutron matter the situation is even worse since  
the isospin dependence of the phenomenological functionals is less constrained.  
The predictive power of such density functionals  at supra-normal  
densities is restricted.   {\it Ab initio}  
calculations predict throughout a soft EOS in the density range  
relevant for heavy ion reactions at intermediate and low  
energies, i.e. up to about three times $\rho_0$.  
There seems to be no way to obtain an  
EOS as stiff as the hard Skyrme force shown in  Fig. \ref{dbeos_fig} or NL3. 
Since the $nn$ scattering lenght is large, neutron matter 
at subnuclear densities is less  
model dependent. The microscopic 
calculations (BHF/DBHF, 
variational) agree well and results are consistent with 
 'exact' Quantum-Monte-Carlo calculations \cite{carlson03}. 
\begin{figure}[h] 
\centerline{
\includegraphics[width=0.8\textwidth]{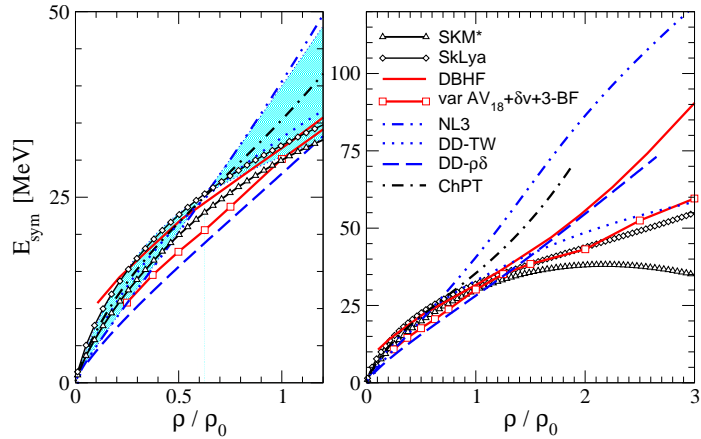}} 
\caption{Symmetry energy as a function of density as predicted by different  
models. The left panel shows the low density region while the right  
panel displays the high density range. 
The Figure is taken from Ref. \protect\cite{WCI}.   
} 
\label{esym_fig} 
\end{figure} 

In isospin asymmetric matter the binding energy is a functional  
of the proton and neutron densities, 
characterized by the  asymmetry parameter $\beta=Y_n-Y_p$ which  
is the difference of the neutron and proton fraction  
$Y_i=\rho_i/\rho~,i=n,p$.  The isospin dependence of the energy   
functional can be expanded in terms of $\beta$ which leads to a  
parabolic dependence on the asymmetry parameter 
\begin{eqnarray} 
E(\rho,\beta) &=& E(\rho) + E_{\rm sym}(\rho) \beta^2 + {\cal O}(\beta^4) +  
\cdots \nonumber \\ 
E_{\rm sym}(\rho) &=& \frac{1}{2}   
\frac{\partial^2E(\rho,\beta)}{\partial \beta^2}|_{\beta=0}  
= a_4 + \frac{p_0}{\rho_{0}^2} (\rho - \rho_0) +\cdots  
\label{esym} 
\end{eqnarray} 
Fig. \ref{esym_fig} compares the symmetry energy predicted from  
the DBHF and variational calculations to that of the  
empirical density functionals already shown in Fig. \ref{nmeos_fig} 
In addition the relativistic DD-$\rho\delta$ RMF functional \cite{baran04}  
is included.  
Two Skyrme functionals, SkM$^*$ and the more recent Skyrme-Lyon force  
SkLya represent non-relativistic models.   
The left panel zooms the low density region while the right panel  
shows the high density behavior of $ E_{\rm sym}$. Remarkable  
is that most empirical models coincide around $\rho \simeq 0.6 \rho_0$  
where   $ E_{\rm sym} \simeq 24$ MeV. This demonstrates that  
constraints from finite nuclei are active for an average density  
slightly above half saturation density. 
However, the extrapolations to supra-normal densities  
diverge dramatically. This is crucial since the high density behavior  
of  $ E_{\rm sym}$ is essential for the structure and the stability of  
neutron stars (see also the discussion in Sec. V.5).  
The microscopic models show a density dependence which can still  
be considered as {\it asy-stiff}. DBHF \cite{dalen04} is thereby stiffer  
than the variational results of Ref. \cite{akmal98}. The density  
 dependence is generally more complex  than in RMF  theory, in  
particular at high densities where $E_{\rm sym}$ shows a non-linear and  
more pronounced increase.   
Fig. \ref{esym_fig} clearly demonstrates the necessity to constrain  
the symmetry energy at supra-normal densities with the help of heavy  
ion reactions. 

The hatched area in Fig. \ref{esym_fig}  
displays the range of $ E_{\rm sym}$ which has  
been obtained by constructing a density dependent  
RMF functional varying thereby the  
 linear asymmetry parameter $a_4$ from 30 to 38 MeV \cite{niksic02}.  
In  Ref. \cite{niksic02} it was concluded that  
charge radii, in particular the skin thickness $r_n - r_p$ in heavy  
nuclei constrains the allowed  
range of  $a_4$ to $32\div 36$ MeV for relativistic functionals. 
 
\subsubsection{Effective nucleon masses} 
The introduction of an effective mass is a common concept to characterize the  
quasi-particle properties of a particle inside a strongly interacting  
medium. In nuclear physics exist different definitions of the  
effective nucleon mass which are  
often compared and sometimes even mixed up:  
the non-relativistic effective mass $m^*_{NR}$ and the  
relativistic Dirac mass $m^*_{D}$. 
These two definitions  are based on different physical concepts.  
The nonrelativistic mass parameterizes 
the momentum dependence of the single-particle potential.  
The relativistic Dirac mass  
is defined through the scalar part of the nucleon self-energy in the  
Dirac field equation which is absorbed into the effective mass  
$m^*_{D} =M + \Sigma_S (k, k_F)$. The Dirac mass is a smooth function of  
the momentum. In contrast, the nonrelativistic effective  
mass - as a model independent result -   
shows a narrow enhancement near the Fermi surface due to an enhanced  
level density \cite{mahaux85}. For a recent review on this subject 
and experimental constraints on  $m^*_{NR}$ see Ref. \cite{lunney03}. 

While the Dirac  mass is a genuine relativistic quantity the effective  
mass  $m^*_{NR}$ is determined by the single-particle energy  
\begin{eqnarray} 
m^*_{NR} = k [dE/dk]^{-1} = \left[\frac{1}{M}  
+  \frac{1}{k} \frac{d}{ dk} U \right]^{-1}~~. 
\label{Landau1} 
\end{eqnarray} 
\begin{figure}[h] 
\centerline{
\includegraphics[width=0.8\textwidth]{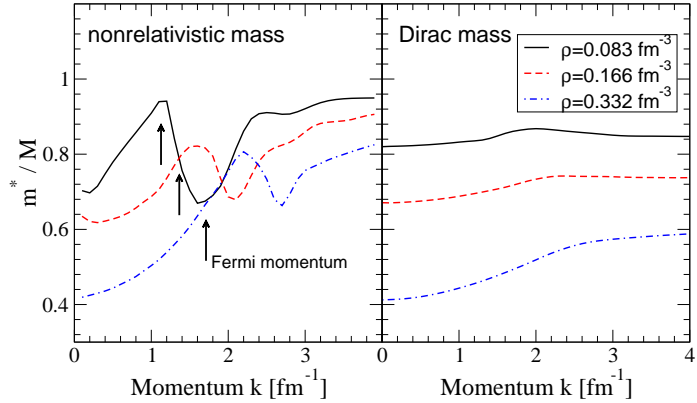}} 
\caption{The effective mass in isospin symmetric nuclear matter  
as a function of the momentum $k$ at different densities  
determined from relativistic Brueckner calculations.   
Figure is taken from Ref. \protect\cite{dalen05}.  
} 
\label{mass1_fig} 
\end{figure} 
$m^*_{NR}$ is a measure of the non-locality of  
the single-particle potential $U$ (real part) which can be due to  
non-localities in space, resulting in  
a momentum dependence, or in time, resulting in an energy dependence.  
In order to clearly separate both effects, one has to distinguish further  
between the so-called k-mass and the E-mass \cite{jaminon}.  
The spatial non-localities of $U$ are mainly  
generated by exchange Fock terms and the resulting k-mass is a smooth  
function of the momentum. Non-localities in time are generated by Brueckner  
ladder correlations due to the scattering to intermediate states which  
are off-shell. These are mainly short-range correlations which generate a  
strong momentum  dependence with a characteristic enhancement of the  
E-mass slightly above the Fermi surface \cite{mahaux85,jaminon,muether04}.  
The effective mass defined  
by Eq. (\ref{Landau1}) contains both, non-localities in  
space and time and is given by the product of k-mass and E-mass \cite{jaminon}.  
In Fig.~\ref{mass1_fig}  
the nonrelativistic effective mass and the Dirac mass, both determined from  
DBHF calculations \cite{dalen05}, are  
shown as a function of momentum $k$ at different Fermi momenta  
of $k_F=1.07,~1.35,~1.7~{\rm fm}^{-1}$.  
$m^*_{NR}$  shows  the  typical peak structure as a function of  
momentum around $k_F$ which is also seen in BHF calculations \cite{muether04}.  
The peak reflects the increase of the  
level density due to the vanishing imaginary part of the optical  
potential at $k_F$ which is also seen, e.g., in shell model calculations  
\cite{mahaux85,jaminon}. One has, however,  
to account for correlations beyond mean field or Hartree-Fock  
in order to reproduce this behavior.  
\begin{figure}[h] 
\centerline{
\includegraphics[width=0.8\textwidth]{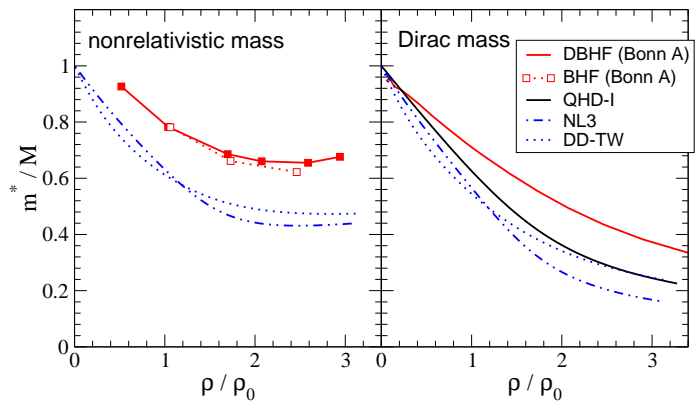}} 
\caption{Nonrelativistic and Dirac effective  
mass in isospin symmetric nuclear matter  
as a function of the density for various models. 
} 
\label{mass2_fig} 
\end{figure} 
Fig.~\ref{mass2_fig} compares the density dependence of the two  
effective masses determined at $k_F$. Both masses decrease with increasing density,  
the  Dirac mass continously, while $m^*_{NR}$ starts to rise  
again at higher densities. Phenomenological density functionals  
(QHD-I, NL3, DD-TW) yield systematically smaller values of  $m^*_{NR}$  
than the microscopic approaches. This reflects the  
lack of nonlocal contributions from short-range and many-body  
correlations in the mean field approaches. 
\subsubsection{Proton-neutron mass splitting} 
A heavily discussed topic is in the moment the proton-neutron mass  
splitting in isospin asymmetric nuclear matter. This question is of  
importance for the forthcoming new generation of radioactive beam  
facilities which are devoted to the investigation of the isospin  
dependence of the nuclear forces at its extremes. However, presently  
the predictions for the  
isospin dependences differ substantially.  
BHF calculations \cite{zuo02,muether04}  
predict a proton-neutron mass splitting of $m^*_{NR,n} > m^*_{NR,p}$.  
This stands in contrast to relativistic mean-field (RMF) theory. When only a  
vector isovector $\rho$-meson is included Dirac phenomenology  
predicts equal masses $m^*_{D,n}= m^*_{D,p}$ while the inclusion of the  
scalar isovector $\delta$-meson, i.e. $\rho+\delta$, leads to  
$m^*_{D,n} < m^*_{D,p}$ \cite{baran04}. When the effective mass is  
derived from RMF theory, it shows the same behavior as the corresponding  
Dirac mass, namely  $m^*_{NR,n} < m^*_{NR,p}$ \cite{baran04}. Conventional  
Skyrme forces, e.g. SkM$^*$, lead to $m^*_{NR,n} < m^*_{NR,p}$ \cite{pearson01} 
while  the more recent Skyrme-Lyon interactions (SkLya) predict the same  
mass splitting as RMF theory.     
The predictions from relativistic DBHF calculations are in the  
literature still controversial. They depend strongly  on approximation schemes and  
techniques used to determine the Lorentz  
and the isovector structure of the nucleon self-energy.  
Projection techniques are involved but more accurate and  
yield the same mass splitting as found in RMF theory when the  
$\delta$ -meson is included, i.e. $m^*_{D,n} < m^*_{D,p}$  
\cite{dalen04,dejong98}. Recently also the non-relativistic  
effective mass has been determined with the DBHF approach and here  
a reversed proton-neutron mass splitting was 
found, i.e. $m^*_{NR,n} >m^*_{NR,p}$  
\cite{dalen05}. Thus DBHF is  in agreement with the results 
from nonrelativistic  
BHF calculations. 
\subsubsection{Optical potentials} 
\begin{figure}[h] 
\centerline{
\includegraphics[width=0.8\textwidth]{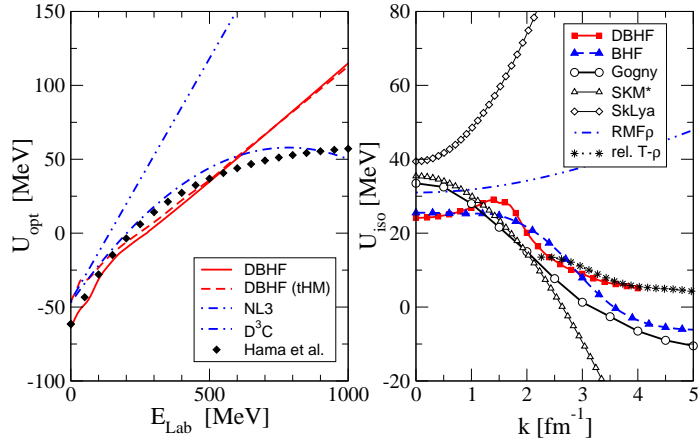}} 
\caption{Nucleon optical potential in nuclear matter at $\rho_0$. 
On the left side DBHF calculations 
for symmetric nuclear matter from
\protect\cite{terhaar87} and \protect\cite{boelting99} are compared to the 
phenomenological models NL3 and D$^3$C \protect\cite{typel05} and to the 
p-A scattering analysis of \protect\cite{hama}. The right panel compares 
the iso-vector optical potential from DBHF \protect\cite{dalen04} and 
BHF   \protect\cite{zuo05} to phenomenological RMF 
\protect\cite{gaitanos04b} , Gogny and 
Skyrme forces and to a relativistic $T-\rho$ 
approximation \protect\cite{trho05}. 
} 
\label{uopt_fig} 
\end{figure}

The second important quantity related to the momentum dependence 
of the mean field is the optical nucleon-nucleus potential. At subnormal 
densities the optical potential $U_{\rm opt}$ is constraint by 
proton-nucleus scattering data \cite{hama} and at supra-normal densities 
constraints can be derived from heavy ion reactions, see Refs.  
\cite{dani00,gaitanos01,giessen2}. In a relativistic framework 
the optical Schroedinger-equivalent 
nucleon potential (real part) is defined as 
\begin{equation}
U_{\rm opt} 
=  - \Sigma_S  + \frac{E}{M} \Sigma_{V} 
        + \frac{\Sigma_S^2  - \Sigma_{V}^2}{2M}~.
\label{uopt}
\end{equation}
One should thereby note that in the literature sometimes also 
an optical potential, given by the difference of the single-particle 
energies in  medium and free space $U= E - \sqrt{M^2 +{\bf k}^2}$ is  
used \cite{dani00} which should be not mixed up with (\ref{uopt}). 
In a relativistic framework momentum independent fields $\Sigma_{S,V}$ 
(as e.g. in RMF theory) 
lead always to a linear energy dependence of $U_{\rm opt}$. 
As seen from Fig.~\ref{uopt_fig} DBHF reproduces 
the empirical optical potential \cite{hama} extracted from 
proton-nucleus scattering for nuclear matter at $\rho_0$ reasonably well 
up to a laboratory energy of about 0.6-0.8 GeV. 
However, the saturating behavior at large momenta cannot 
be reproduced by this calculations because of missing 
inelasticities, 
i.e. the excitation of isobar resonances above the pion threshold. When such continuum 
excitations are accounted for optical model caculations are 
able to describe nucleon-nucleus scattering data also at 
higher energies \cite{geramb02}. In heavy ion 
reactions at incident energies above 1 AGeV such a saturating behavior is 
required in order to reproduce transverse flow observables \cite{giessen2}.
One has then to rely on phenomenological approaches where 
the strength of the vector potential is artificially suppressed, e.g. by 
the introduction of additional form factors \cite{giessen2} or by 
energy dependent terms in the QHD Lagrangian \cite{typel05} (D$^3$C model 
in Fig.\ref{uopt_fig}) . 

The isospin dependence, expressed by the 
isovector optical potential $U_{\rm iso}= (U_{\rm opt,n} - U_{\rm opt,p}) / (2
\beta)$ is much less constrained by data. The knowledge of 
this quantity is, however, 
of high importance for the forthcoming radioactive beam experiments.  
The right panel of Fig. \ref{uopt_fig} compares the predictions 
from DBHF \cite{dalen04} and BHF \cite{zuo05} to the phenomenological 
Gogny and Skyrme (SkM$^*$ and SkLya) forces and a relativistic $T-\rho$ 
approximation \cite{trho05} based on empirical NN scattering 
amplitudes \cite{neil83}. At large momenta 
DBHF agrees with the tree-level results of Ref. \cite{trho05}.
While the dependence of  $U_{\rm iso}$ on the 
asymmetry parameter $\beta$ is found to be rather weak \cite{dalen04,zuo05}, 
the predicted energy and density dependences are quite 
different,  in particular between  the microscopic and the 
phenomenological approaches. The energy dependence 
of $U_{\rm iso}$ is very little constrained by data. The old 
analysis of optical potentials of scattering on charge asymmetric 
targets by Lane \cite{lane62} is consistent with a decreasing 
potential as predicted by DBHF/BHF, while more recent analyses 
based on Dirac phenomenology \cite{madland89} come to the opposite 
conclusions. RMF models show a linearly increasing
 energy dependence of $U_{\rm iso}$ (i.e. quadratic in $k$)  
like SkLya, however 
generally with a smaller slope (see discussion in Ref. \cite{baran04}). 
To clarify this question certainly more experimental 
efforts are necessary.  
\subsection{Probing the EOS by kaon production in heavy ion reactions}
With the start of the first relativistic heavy ion programs the 
hope was that particle production would provide a direct experimental 
access to the nuclear EOS \cite{stock86}. It was expected that the  
compressional energy should be released into the creation of 
new particles, primarily pions, when the matter expands \cite{stock86}. 
However, 
pions have large absorption cross sections and they turned out not to 
be suitable messengers of the compression phase. They undergo several 
absorption cycles through nucleon  resonances
and freeze out at final stages of the reaction 
and at low densities. Hence pions loose most of their knowledge on the 
compression phase and are not very sensitive probes for 
stiffness of the EOS. 

\begin{figure}[h]
\centerline{
\includegraphics[width=0.6\textwidth]{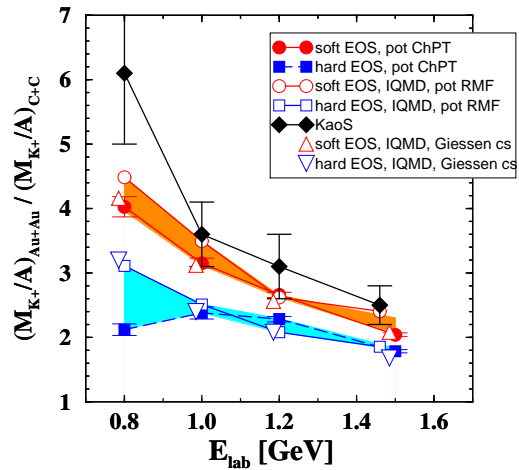}} 
\caption{Excitation function of the ratio $R$ of $K^+$ 
multiplicities obtained in inclusive Au+Au over C+C 
reactions. RQMD \protect\cite{fuchs01} and IQMD 
\protect\cite{hartnack05} calculations are compared 
to KaoS data \protect\cite{sturm01}. Figure is taken from 
\protect\cite{fuchs05}.
}
\label{fig_ratio_1}
\end{figure}

After pions turned out to fail as suitable messengers, 
$K^+$ mesons were suggested as promising tools to probe the nuclear 
EOS \cite{AiKo85}. At subthreshold energies 
$K^+$ mesons are produced in the high density phase and due to 
the absence of absorption reactions they have a long mean free path 
and leave the matter undistorted by strong final state interactions. 
Moreover, at subthreshold energies nucleons have to accumulate energy by 
multiple scattering processes in order to overcome the threshold 
for kaon production and therefore these processes should be particularly 
sensitive to collective effects. 
Within the last decade the KaoS Collaboration has performed systematic 
measurements of the $ K^+$ production far below threshold, see Refs.  
\cite{kaos94,barth,kaos99,laue00,sturm01}. Based on the new data situation, 
in Ref. \cite{fuchs01} the question 
if valuable information on the nuclear EOS can be extracted 
has been revisited and it has been 
shown that subthreshold  $K^+$ production provides indeed a suitable and 
reliable tool for this purpose. These results have been confirmed by 
the Nantes group later on \cite{hartnack01}. In subsequent investigations 
the stability of the EOS dependence has been proven, Refs.  
\cite{fuchs01b,fuchs05,hartnack05}. 
\begin{figure}[h]
\centerline{
\includegraphics[width=0.8\textwidth]{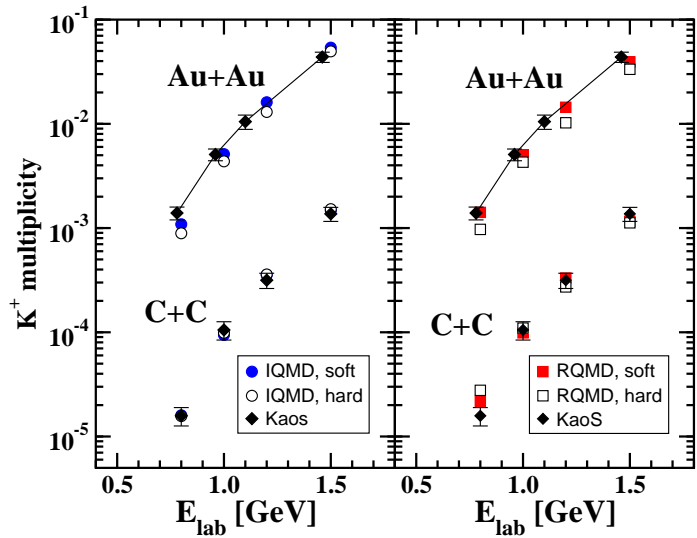}} 
\caption{Excitation function of the $K^+$ multiplicities 
in $Au+Au$ and $C+C$ reactions. RQMD \protect\cite{fuchs01} and 
IQMD \protect\cite{hartnack05} with 
in-medium kaon potential and using a hard/soft nuclear EOS
are compared to data from the KaoS Collaboration \protect\cite{sturm01}. 
}
\label{fig_ex_2}
\end{figure}
Excitation functions from KaoS \cite{sturm01,kaos99} are 
 shown in Fig. \ref{fig_ex_2} and compared to RQMD \cite{fuchs01,fuchs05} 
and IQMD \cite{hartnack05} calculations. As expected the EOS dependence 
is pronounced in the Au+Au system while the light C+C system 
serves as a calibration. 
The effects become even more evident when the ratio $R$ of the 
kaon multiplicities obtained in Au+Au over C+C 
reactions (normalised to the corresponding mass numbers) is built 
\cite{fuchs01,sturm01}. Such a ratio has the advantage that 
possible uncertainties which 
might still exist in the theoretical calculations should cancel out 
to large extent. This ratio is shown in Fig. \ref{fig_ratio_1}. 
Both, soft and hard EOS, 
show an increase of $R$ with decreasing energy 
down to 1.0 AGeV. However, this increase is much less 
pronounced when the stiff EOS is employed. The comparison to the experimental 
data from KaoS \cite{sturm01}, where the increase of $R$ is even more 
pronounced, strongly favours a soft equation of state. 
Fig. \ref{fig_ratio_1} demonstrates also the robustness of this observable. 
Exploring the range of uncertainty in the corresponding transport calculations 
the stability of the conclusions drawn from this observable has been 
demonstrated in Ref. \cite{hartnack05}. This concerns elementary input,  
in particular the elementary production cross sections 
$N\Delta; \Delta\Delta \mapsto NYK^+$ which are not constrained by data. 
\subsection{Constraints from neutron stars}
Measurements of ``extreme'' values,
like large masses or radii, huge luminosities etc.\
as provided by compact stars offer good opportunities to gain deeper insight
into the physics
of matter under extreme conditions. 
There has been substantial progress in recent time from the astrophysical 
side. 

The most spectacular observation was probably the 
recent measurement \cite{NiSp05} on PSR J0751+1807, a millisecond pulsar in a binary 
system with a helium white dwarf secondary, which implies
a pulsar mass of $2.1\pm0.2\left(^{+0.4}_{-0.5}\right) {\rm M_\odot}$
with $1\sigma$ ($2\sigma$) confidence. 
Therefore, a reliable EOS has to describe neutron star
(NS) masses of at least $1.9~ {\rm M}_\odot$ ($1\sigma$) in a strong,
or  $1.6~ {\rm M}_\odot$ ($2\sigma$) in a weak interpretation.
This  condition limits the softness of EOS in NS matter. One might 
therefore be worried about an apparent contradiction between the 
constraints derived from neutron stars and those from heavy ion 
reactions. While heavy ion reactions favor a soft EOS, PSR J0751+1807  
requires a stiff EOS. The 
corresponding constraints are, however, 
complementary rather than contradictory. 
Intermediate energy heavy-ion reactions, e.g. subthreshold kaon production, 
constrains the EOS at densities up to $2\div3~\rho_0$ while the maximum NS 
mass is more sensitive to the high density behaviour of the EOS. Combining the 
two constraints implies that the EOS should be {\it soft at moderate 
densities and  stiff at high densities}. Such a behaviour is predicted 
by microscopic many-body calculations (see Fig.~\ref{dbeos_fig}). DBHF, BHF 
or variational calculations, typically, lead to maximum NS masses between  
$2.1\div 2.3~M_\odot$ and are therefore in accordance with PSR J0751+1807, 
see Ref. \cite{klaehn06}. 

There exist several other constraints on the nuclear EOS which can 
be derived from observations of  compact stars, see e.g. 
Refs. \cite{klaehn06,steiner05,steiner05b}. 
Among these, the most promising one is the Direct Urca (DU) process which 
is essentially driven by the proton fraction inside the NS \cite{lattimer91}. 
DU processes, e.g. the neutron $\beta$-decay $n\to p+e^-+\bar\nu_e$,
are very efficient regarding their neutrino production,
even in superfluid NM ~\cite{BlGrVo04,KoVo05},
and cool NSs too fast to be in accordance
with data from thermally observable NSs.
Therefore, one can suppose that
no DU processes should occur
below the upper mass limit
for ``typical'' NSs, i.e.
$M_{DU}\geq 1.5~M_\odot$ ($1.35~M_\odot$ in a weak interpretation).
These limits come from a population synthesis of young,
nearby NSs~\cite{Popov:2004ey} and masses of NS binaries  ~\cite{NiSp05}.

\section{Summary}
The status of theoretical models which make predictions for the EOS 
can roughly be summarized as follows: phenomenological density 
functionals such as Skyrme, Gogny or relativistic mean field models 
provide high precision fits to the nuclear chart but extrapolations 
to supra-normal densities or the limits of stability are highly 
uncertain. A more controlled way provide effective field theory 
approaches which became quite popular in recent time. Effective 
chiral field theory allows e.g. a systematic generation of two- 
and many-body nuclear forces. However, these approaches are low 
momentum expansions and when applied to the nuclear many-body 
problem, low density expansions. Ab initio calculations for the 
  nuclear many-body problem such as variational or Brueckner 
calculations have reached a high degree of sophistication and 
can serve as guidelines for the extrapolation to the regimes of 
high density and/or large isospin asymmetry. Possible future 
devellopments are to base such calculations on modern EFT 
potentials and to achieve a more consistent treatment of 
two- and three-body forces. 

If one intends to constrain these models by nuclear reactions 
one has to account for the reaction dynamics by semi-classical 
transport models of a Boltzmann or molecular dynamics type. 
Suitable observables which have been found to be sensitive on 
the nuclear EOS are directed and elliptic 
collective flow pattern and particle production, 
in particular kaon production, at higher energies. Heavy ion 
data suggest that the EOS of symmetric nuclear matter shows a soft 
behavior in the density regime between one to about three times 
nuclear saturation density, which is consistent with the 
predictions from many-body calculations.  Conclusions 
on the EOS are, however, complicated by the interplay between the 
density and the momentum dependence of the nuclear mean field. 
Data which constrain the isospin dependence of the mean field are 
still scare. Promising observables are isospin diffusion, iso-scaling 
of intermediate mass fragments and particle ratios 
($\pi^+/\pi^-$ and eventually $K^+/K^0 $ \cite{theo06}). Here the situation 
will certainly improve when the forthcoming radioactive beam 
facilities will be operating.

\bibliographystyle{ws-procs9x6}

\end{document}